\documentclass[12pt,a4paper]{article}
\usepackage{amsmath}
\usepackage[dvips]{graphicx}
\input epsf
\usepackage{times}
\begin{document}
\begin{titlepage}
\title{Impact parameter model for  GPD}
\author{S.M. Troshin, N.E. Tyurin\\[1ex]
\small  \it Institute for High Energy Physics,\\
\small  \it Protvino, Moscow Region, 142281, Russia} \normalsize
\date{}
\maketitle

\begin{abstract}
We propose a dynamical model for the unpolarized generalized parton distribution function (GPD)
based on impact parameter representation and discuss its   $x$ and $t$ dependencies
at large and small values of these variables.
\end{abstract}
\end{titlepage}
\setcounter{page}{2}
  Recent developments in the  nucleon structure studies are
closely related to
 the generalized parton distributions (GPDs) $H_q(x,\xi,t)$ and others.
 These functions are the matrix elements
 of the quark field operators between nucleon states of the different momenta with $x=1/2(x_f+x_i)$ and
 $\xi=1/2(x_f-x_i)$\footnote{Note, that $x_i$ and $x_f$ are the
light-cone momentum fractions of the quark in the initial and
final state, respectively.}.
 This topic attracted much attention
 in the last decade. The results
 are collected in the  review papers  \cite{gok,dil,bur,ji,rad,feld,bof}
 and many  references to the original papers can be found therein.
The important point  is  that the Fourier transform of the
function $H_q(x,\xi,t)$ can be related to a weighted parton number
density corresponding to the quark $q$ flavor  in the transverse
plane when interpolated to the point $\xi=0$ \cite{bur} where the variable $\xi$ is
  the longitudinal momentum transfer ( skewedness parameter).
The impact parameter-dependent GPD allows one to obtain
 information on the parton distribution in the
transverse plane over the nucleon area simultaneously with the parton
distribution over fraction $x$ of the total nucleon momentum. This
result would pave the roads from  the experimental data related to
the amplitudes of the processes $\gamma^*p\to\gamma p$ (deeply
virtual Compton scattering
 (DVCS)) and hard meson
exclusive production $\gamma^*p\to M p$  to the studies of hadron structure
in the two-dimensional transverse space.   For the first time position space
 aspects of  parton model were studied  in the
paper \cite{soper}.

A common procedure of obtaining GPD dependence is to  guess the
form of $H_q(x,\xi,t)$
 proceeding from the model assumptions, namely use of the general constraints based on
 form-factors, first principles, experience gained from high-energy
 scattering and experimental studies of DVCS \cite{cudell,gdl,dil1,jenk,ter,ahm, kum}.
 However, since impact parameter representation  provides  a useful framework for
 implementing geometrical  aspects
 as well as transparent account of the unitarity constraints, it can be used as
  a complementary way for the GPD's modelling.  In particular, unitarity of the
  quark-hadron scattering amplitude at low values of $x$
 can be  implemented in that way \cite{lowx} and forms of
 gluon distributions were elaborated  on the
 base of this representation in \cite{str,bond}. Impact parameter
 representation is also very helpful in establishing the relation of the GPD dependence
with color transparency phenomena \cite{mill,burk,liu}.
In this note we continue to proceed that way and  construct a model for the impact
parameter dependent GPD to provide further insight into the spatial hadron properties.

Prior to the model description we would like to provide several
known general relations for GPDs. We  consider generalized parton
distribution function $H_q(x,\xi,t)$, i.e. the case of unpolarized
hadron and parton. In the forward limit this function reproduces
standard, integrated over impact parameter, parton distribution
function, $H_q(x,0,0)=q(x)$. If one uses the combination
\[
{\cal{H}}_q(x,t)=H_q(x,\xi=0,t)+H_q(-x,\xi=0,t),
\]
then the  nucleon Dirac form factor $F_1(t)$ is the sum $F_1(t)\equiv \sum_q e_qF_1^q(t)$
where  $F_1^q(t)$ is represented as the
integral $F_1^q(t)=\int_0^1 {\cal{H}}_q(x,t)$. Relation of the
impact parameter dependent parton distribution with the
corresponding GPD was established in \cite{bur0}
 and  has the
form
\begin{equation}\label{fb}
q(x,b)=\frac{1}{2\pi}\int_0^\infty
\sqrt{-t}d\sqrt{-t}{\cal{H}}_q(x,t)\mbox{J}_0(b\sqrt{-t}),
\end{equation}
where $b=|\mathbf{b}|$ is the distance from the active quark
position to the hadron center of mass in the transverse plane,
which is determined by the relation
\begin{equation}\label{cms}
\sum_ix_i \mathbf{b}_i=0,
\end{equation}
and $x_i$ stands for the longitudinal momentum fraction of the
parton $i$ in the infinite momentum frame. It should be noted that
the function $q(x,b)$, in contrast to the function $q(x)$, is dimensional,
 it has dimension of a
squared mass and is interpreted as a parton number density referred to the
transverse area of a nucleon.
  It can be written therefore in the form $q(x,b)=\tilde{\mu}^2\tilde{q}(x,b)$,
where $\tilde{\mu}$ is a some mass scale related, as it will be clarified later,
 to the finite transverse size of a parton
 and $\tilde{q}(x,b)$ is a dimensionless parton number
density distribution, i.e. it can be interpreted as
a probability to find  quark  of  flavor $q$ with the the momentum
fraction $x$ of the total nucleon momentum at the distance $b$ from the hadron's center.

Another important parameter is the distance from the
active quark to the center of mass of other constituents
(spectators)
\begin{equation}\label{dis}
\mathbf{r}_\perp=\frac{\mathbf{b}}{1-x}.
\end{equation}
Its mean value can also be associated with the hadron size.
Requirement that the value $\langle \mathbf{r}_\perp^2\rangle _x $
remains finite at $x\to 1$ leads to $t$-dependence of GPD in the
form of the product $(1-x)^n t$, $n\geq 2$ in the limit $x\to 1$
\cite{burk}. Another  requirement is that the values of $\langle \mathbf{b}^2
\rangle _x $ and, of course, $\langle \mathbf{r}_\perp^2\rangle _x
$, should not increase faster than $\ln^2{\frac{1}{x}}$ at $x\to
0$. It has grounds in the short-range nature of strong
interactions and unitarity.

Keeping in mind the above preconditions, we
 propose the following form  for the  function $\tilde{q}(x,b)$:
\begin{equation}\label{qsb}
\tilde{q}(x,b)=\frac{\tilde{q}_0(x,b)}{1+\tilde{q}_0(x,b)}.
\end{equation}
Such form for
$\tilde{q}(x,b)$ can be obtained with  account for the finite size of the parton in the transverse area
( determined by the inverse mass
$1/\tilde{\mu}$) in a way similar to the one used in \cite{cleymans} for the case of the gas of extended
hadrons \footnote{Indeed, the number density of
point-like partons $n_0$ can be written in the form
$n_0=\frac{N}{(S_h-S_qN)}$,
where $N$ is the total number or partons (located in the hadron transverse area $S_h$)
with transverse size $S_q$.}.
The form of Eq. \ref{qsb} can  also be related to
 the form of the quark-hadron
amplitude as it was discussed in \cite{lowx}.

The function $\tilde{q}_0(x,b)$ is a dimensionless
distribution function for the point-like partons. It should be noted
that
$\tilde{q}(x,b)\simeq \tilde{q}_0(x,b)$ in the region where $\tilde{q}_0(x,b)\ll
1$. It takes place, e.g. at large values of $b$,
where the finite size of partons does not play a role.

Construction of the function $q(x,b)$ in the form of Eq. \ref{qsb}
 implies that the weighted parton number density referred to the
  transverse area of the nucleon cannot exceed  the magnitude of inverse parton transverse area.
This form and the model for the function $\tilde{q}_0(x,b)$ described below lead to the saturation of the
 parton distribution function $q(x,b)$ at low values of $x$ and $b$. The mass scale $\tilde{\mu}$
 determines  transverse parton size
 and can be related to the constituent quark radius.  We note also,
 that the following upper bound takes place:
\[
 q(x,b)\leq 1/S_q,
 \]
 where $S_q$ is a transverse area of an extended parton.  Indeed, there are experimental
 indications on the presence of the extended objects inside the proton found in the DIS data
 \cite{petron}.

To construct the function $\tilde{q}_0(x,b)$, a dimensionless distribution of the point-like
partons, we consider the picture of hadron structure described in \cite{diak}.
In this approach hadrons contain massive quarks interacting with self-consistent pion field and the
pions  themselves are not elementary.
They consist of strongly bounded quark-antiquark pairs.
 This picture is a model-independent
consequence of the spontaneous symmetry breaking. Thus, the active quark moving in the field
of the spectators interacts with them by a pion exchange. This is strong  interaction and it cannot be
reduced therefore to one-pion exchange, and  spectators should not be considered as
the remaining valence quarks only. Being essentially relativistic,
the bound state problem should be addressed on
the multiparticle basis and Fock components with many partons should be taken into account.
Thus, the dynamics of the quark interactions and probability to
find quark at the distance $b$  from the hadron center  depend, in fact,
on the distances between active and spectator quarks.  After averaging over positions of the spectator
quarks, the  $r_\perp = |\mathbf{r}_\perp|$ will be the only distance the  function $\tilde{q}_0(x,b)$
should depend on.
The  described  mechanism  implies that the lifetimes of the active and spectator quarks are comparable,
i.e. the process is coherent. Such a coherence can occur even in the case when an active
quark is hard and has large value of $x$ and spectator quarks are soft and have small values of $x$.
The interaction of the active quark with spectator ones should be considered as a final-state process
as it was discussed in \cite{hoyer}.

For the input distribution function $\tilde{q}_0(x,b)$ we use a factorized   form
\[
\tilde{q}_0(x,b)=\chi (x)\rho (r_\perp).
\]
With account for quark interactions with the pion field a simple exponential
dependence was chosen for the function $\rho (r_\perp)$,
\begin{equation}\label{ro}
\rho (r_\perp)\sim \exp (-\mu r_\perp),
\end{equation}
where mass scale $\mu$ is order of the pion mass (but should not be exactly equal to it).
Of course, this is oversimplified form because of
 unspecified pre-exponential factors dependent on the variable
 $r_\perp$ can be presented also.
 For the purposes of  qualitative consideration we  neglect
these complications. The function $\chi(x)$
can be connected somehow to a number of the spectator quarks interacting with active
 quark (having the fraction of nucleon momentum equal to $x$).

 For the behavior of $\chi (x)$ in the limiting
 cases $x\to 0$ and $x\to 1$, we use the standard
dependencies borrowed from the  deep-inelastic processes studies of the parton distribution functions:
\begin{equation}\label{xto01}
\chi (x) \sim x^{-\lambda} \quad \mbox{and} \quad \chi (x) \sim (1-x)^N
\end{equation}
respectively. Note, that  parameter $\lambda > 0$.
Thus, the adopted expression for the function $\tilde{q}_0(x,b)$ is the following
\begin{equation}\label{qosb}
\tilde{q}_0(x,b)=\chi (x)\exp (-\frac{\mu b}{1-x}).
\end{equation}

The function $\chi (x)$ can, in principle, be obtained from the parton distribution  $q(x)$
with the use of Eq. (\ref{qsb}), i.e. due to relation
\begin{equation}\label{li2}
q(x)\sim {-(1-x)^2}\mbox{Li}_2[-\chi(x)],
\end{equation}
where the function $\mbox{Li}_2(z)$ is dilogarithm\footnote{The dilogarithm is defined by the series
$\mbox{Li}_2(z)=\sum_{k=1}^\infty \frac{z^k}{k^2}$ and at $|z|\geq 1$ this function defined through an analytic
continuation.}.
Qualitative behavior of the dimensionless parton distribution $\tilde{q}(x,b)$
 with the function $\chi (x)$ having limiting dependencies in the form of Eq.
(\ref{xto01}) is illustrated by the plot in Fig.~1. Evidently,  the
function $\tilde{q}(x,b)$ is saturated in the region of small $x$ and $b$.
\begin{figure}[h]
\begin{center}
  \resizebox{8cm}{!}{\includegraphics*{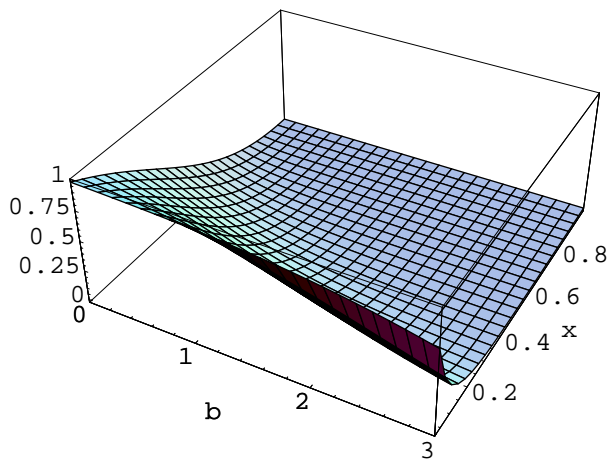}}
\end{center}
\caption{Dimensionless generalized parton distribution function  $\tilde{q}(x,b)$ (color online).}
\end{figure}

The average value $\langle \mathbf{b}^2\rangle _x$ is determined by the relation
\begin{equation}\label{aver}
\langle \mathbf{b}^2\rangle _x =\frac{\int_0^\infty \mathbf{b}^2 q(x,b) bdb}
{\int_0^\infty  q(x,b) bdb}
\end{equation}
and
\[
\langle \mathbf{r}_\perp^2\rangle _x =\frac{\langle \mathbf{b}^2\rangle _x}{(1-x)^2}.
\]
It is clear that  $q(x,b)$ at large values of $x$
has the same dependence as the function $\tilde{q_0(x,b)}$
 and it  immediately
provides at $x\to 1$ the finite value for $\langle
\mathbf{r}_\perp^2\rangle _x $, i.e.
\[
\langle \mathbf{r}_\perp^2\rangle _x \sim 1/\mu^2,\quad\mbox{at}\quad x\to 1
\]
while
\[
\langle \mathbf{b}^2\rangle _x \sim (1-x)^2/\mu^2.
\]

At $x\to 0$  the expression for the function $q(x,b)$ in the
form of Eq. (\ref{qsb}) with limiting dependence of Eq. (\ref{xto01}) provides
\[
\langle \mathbf{b}^2\rangle _x \simeq \langle \mathbf{r}_\perp^2\rangle _x \sim \frac{\lambda^2}{\mu^2}\ln^2 (\frac{1}{x}).
\]

Thus, the model leads to reasonable  results for the hadronic sizes at small and large
values of $x$. At large values of $x$ the  hadron size is $x$-independent. It has finite value and
 consistent  with quark confinement.
At  small values of $x$ hadronic size increases  in a way consistent with unitarity
and  short-range nature of strong interactions.
Such behavior is a result of transform of the $x$-independent $b$-dependence of $\tilde{q}_0(x,b)$
(at small values of $x$) into the function $q(x,b)$ which results in  average squared
impact parameter
\[
\langle \mathbf{b}^2\rangle _x=4\frac{\partial}{\partial t}\ln {\cal{H}}(x,t)|_{t=0},
\]
rising logarithmically with $x\to 0$, i.e.  $\sqrt{\langle \mathbf{b}^2\rangle _x}\sim\ln (\frac{1}{x})$.
 One can conclude therefore  that hadrons participating in hard interaction have
  finite radius (and this does not imply color transparency phenomenon),
  while soft hadron interaction radius is increasing with energy
  like $\ln s$ (since $s\sim 1/x$).

We consider now the  $x$ and $t$-dependencies of the function ${\cal{H}}(x,t)$
 at large values of $x$:
\begin{equation}\label{hxt}
{\cal{H}}_q(x,t)\sim
\frac{(1-x)^2\chi(x)}{[1-(1-x)^2t/\mu^2]^{3/2}}.
\end{equation}
This  form is  the Fourier-Bessel transform of the function
$q(s,b)$. From Eq. (\ref{hxt}) it is evident that at fixed values
of $-t$ the function ${\cal{H}}_q(x,t)$ does not depend on this
variable in the limit $x\to 1$. If we take the fixed values of $x$
and large values of $-t$ the function ${\cal{H}}_q(x,t)$ seems to
decrease as $(-t)^{-3/2}$. But it is not legitimate to consider
the above two limits independently,
 separating $x$ and $t$ dependencies, because those limits are
  not consistent simultaneously with the requirement for the
parton process to be coherent, which  is a precondition for the validity of the discussed mechanism.
Instead,  we should impose additional
relations which guarantee coherence  at large values of $-t$, e.g.
\begin{equation}\label{bb}
(1-x)t=const.
\end{equation}
or
\begin{equation}\label{bb1}
(1-x)\sqrt{-t}=const.
\end{equation}
 Eq. (\ref{bb}) is known as the Berger-Brodsky limit \cite{bb} and it provides the condition for the
process of parton interactions remain  coherent as well as Eq. (\ref{bb1}) does (cf. \cite{hoyer}).
Thus, when we consider asymptotical dependence on $-t$ we should always take $x\to 1$, otherwise coherence
would be destroyed.
Asymptotical behavior of form-factor, which is an integral
\[
F_1^q(t)=\int_0^1 dx {\cal{H}}_q(x,t),
\]
  has the following power-like form :
 \[
 F_1^q(t)\sim \left(-\frac{\mu^2}{t}\right)^{\frac{N+3}{2}}.
\]

In the region of small values of $x$ the function ${\cal{H}}_q(x,t)$ has the following
dependence\footnote{It results from behavior  of the Bessel function $\mbox{J}_1(x)$
and $\chi(x)$ at small values of $x$.}
at small $-t$:
\begin{equation}\label{lxt}
{\cal{H}}_q(x,t)\sim \ln^2 \left( \frac{1}{x}\right)\exp \left[\frac{\lambda^2}
{8\mu^2}t\ln^2\left(\frac{1}{x}\right)\right].
\end{equation}
In the framework of the adopted picture for the hadron structure  this low-$x$ dependence
should be associated with the sea quarks. Eq. (\ref{lxt}) is in  a qualitative agreement
with the experimentally observed exponential $t$-dependence of the  differential cross-section
of the DVCS process \cite{aaron}.

Thus, we can conclude that the GPD is no constrained by the form factor
 alone, which
is sensitive to the  difference of quark-antiquark number densities only. Instead,
in addition to the amplitudes of the  DVCS processes, it would be promising
to relate GPD behavior to  the amplitudes of another exclusive process
 such as elastic hadron-hadron scattering and use experimentally
measured differential
cross-section of elastic scattering to constrain the $t$-dependence of GPDs.
Large angle elastic scattering is sensitive
to the collisions with small impact parameter and
 provides information on the hadron structure in the
 central region in the position space.

Another  possibility
is related to the multiparticle production. Proceeding along the lines described in \cite{chang}, it is possible
 to treat the amplitude of such process
\[F(\mathbf{B}_1-\mathbf{B}_2,\mathbf{b}_1,x_1....\mathbf{b}_n,x_n, \mathbf{b}'_1,x'_1....\mathbf{b}'_{n'},x'_{n'})\]
as the wave function of the production of the two clusters with $n$ and $n'$ particles with centers
located in the position space at $\mathbf{B}_1=\sum_{i=1}^n x_i\mathbf{b}_i$
and $\mathbf{B}_2=\sum_{i=1}^{n'} x'_i\mathbf{b'}_i$.
Using idea of quark-hadron duality, one can try then to extract information on the GPDs from the observables
measured in the hadron production
processes. This possibility will be pursued  elsewhere. Here we note  that the
 hard inclusive
processes are  complementary to hard elastic scattering since they are sensitive not only to the collisions
 with small impact parameters, but to the peripheral collisions also \cite{dpe}.

To conclude, we   note that the  proposed model for the impact parameter-dependent GPD
  allows one to consider simultaneously the regions  of large
and small values of $x$ and $-t$.  The parton distribution $q(x,b)$ is saturated at
small values of $x$ and $b$.
It provides finite value for the hadron size determined  by the inverse pion mass at large
values of $x$ and logarithmically growing  hadron size at small values of $x$.
 This growth is due to  saturation of the weighted probability
distribution $q(x,b)$ in the impact parameter representation.

We are grateful to A. Kisselev, V. Petrov and B. Pire for the interesting discussions and useful
 remarks on the list of references.

\end{document}